\documentclass[twocolumn,preprintnumbers,amsmath,amssymb]
{revtex4-2}
\usepackage{graphicx}% Include figure files
\usepackage[normalem]{ulem}
\usepackage[svgnames]{xcolor}
\usepackage{bm}% bold math
\usepackage{multirow}\usepackage{float}
\usepackage{titlesec}
\usepackage{soul}
\usepackage{xcolor}
\usepackage{subfigure}
\usepackage{subfig}
\usepackage{lipsum, babel}
\usepackage[font=small,labelfont=bf,
   justification=justified,
   format=plain]{caption}  
\usepackage{multirow}
\usepackage[normalem]{ulem}

\begin{document}
\title{Structural investigation of the quasi-one-dimensional topological insulator Bi$_4$I$_4$}

\author{C. David Hinostroza$^{1}$}
\author{Leandro Rodrigues de Faria$^{3}$}
\author{Gustavo H. Cassemiro$^{2}$}
\author{J. Larrea Jiménez$^{1}$}
\author{Antonio Jefferson da Silva Machado$^{3}$}
\author{Walber H. Brito$^{2}$}
\author{Valentina Martelli$^{1}$}
\thanks{}

\affiliation{(1) Laboratory for Quantum Matter under Extreme Conditions,  Instituto de Física, Universidade de São Paulo, 05508-090, São Paulo, Brazil\\
(2) Departamento de F\'isica, Universidade Federal de Minas Gerais,  C. P. 702, 30123-970, Belo Horizonte, MG, Brazil\\
(3) Escola de Engenharia de Lorena - DEMAR, Universidade de Sao Paulo, 12612-550, Lorena, Brazil\\}

\date{\today} 
\begin{abstract}

The bismuth-halide Bi$_4$I$_4$ undergoes a structural transition around $T_P\sim 300$K, which separates a high-temperature $\beta$ phase ($T>T_P$) from a low-temperature $\alpha$ phase ($T<T_P$). $\alpha$ and $\beta$ phases are suggested to host electronic band structures with distinct topological classifications. Rapid quenching was reported to stabilize a metastable $\beta$-Bi$_4$I$_4$ at $T<T_P$, making possible a comparative study of the physical properties of the two phases in the same low-temperature range. 
In this work, we present a structural investigation of the Bi$_4$I$_4$ before and after quenching together with electrical resistivity measurements.
We found that rapid cooling does not consistently lead to a metastable $\beta$-Bi$_4$I$_4$, and, a quick transition to $\alpha$-Bi$_4$I$_4$ is observed. As a result, the comparison of putative signatures of different topologies attributed to a specific structural phase should be carefully considered. 
The observed phase instability is accompanied by an increase in iodine vacancies and by a change in the temperature dependence of electrical resistivity, pointing to native defects as a possible origin of our finding. Density functional theory (DFT) calculations support the scenario that iodine vacancies, together with bismuth antisites and interstitial are among the defects that are more likely to occur in Bi$_4$I$_4$ during the growth.
\end{abstract}
\maketitle

%%%%%%%%%%%%%%%%%%%%%%%%%%%%%%%%%%%%%%%%%%%%%%%%%%%%%%%%%%%%
%%%%%%%%%%%%%%%%%%%%   INTRODUCTION    %%%%%%%%%%%%%%%%%%%%%
%%%%%%%%%%%%%%%%%%%%%%%%%%%%%%%%%%%%%%%%%%%%%%%%%%%%%%%%%%%%

\section{INTRODUCTION}

Bi$_4$I$_4$ is a quasi-one-dimensional material built by a stack of molecular chains joined together by Van der Waals interactions \cite{von1978bismuth,huang_quasi-1d_2016}. The difference in stacking order gives rise to two different structural phases, dubbed $\alpha$-Bi$_4$I$_4$ and $\beta$-Bi$_4$I$_4$. %(Fig. \ref{fig:cells} (a) and (b), respectively).
The structural transition from $\alpha$ to $\beta$-Bi$_4$I$_4$ is observed crossing a temperature around $T_P\sim 300$  K \cite{huang2021room} and the role that each structure plays in the determination of the topological classification of the electronic bands is still a matter of debate \cite{liu2016weak,noguchi_weak_2019,autes_novel_2016}.
%\walber{The intricate connection between the crystalline structures and their topological properties makes this system an ideal platform for studying topological phase transitions induced by temperature. The possibility of the coexistence of distinct topological phases at low temperatures can also be found in the Sb$_2$Te$_3$ \cite{Serebryanaya2015}, although the pressure required is 4 GPa. In a similar quasi-one dimensional compound, Bi$_4$Br$_4$, previous studies have pointed out to a possible coexistence of a non-trivial topological phase with superconductivity~\cite{Li2019}. The authors also emphasized the importance of the structural phase transitions under high pressure and the emergence of distinct quantum phases.}
The possibility of the coexistence of distinct topological phases can also be found in the Sb$_2$Te$_3$ \cite{Serebryanaya2015}; however, the pressure required is 4 GPa to stabilize into the metastable form.
In a similar quasi-one dimensional compound, Bi$_4$Br$_4$, previous studies have pointed out to a possible coexistence of a non-trivial topological phase with superconductivity~\cite{Li2019}. The authors also emphasized the importance of the structural phase transitions under high pressure and the emergence of distinct quantum phases. In Bi$_4$I$_4$, the intricate connection between the crystalline structures readily achievable at ambient pressure and their topological properties makes this system ideal for studying topological phase transitions induced by temperature.
%%%The related compound and higher order topological insulator Bi$_4$Br$_4$, \cite{noguchi2021evidence} only shows one stable phase at ambient pressure \cite{Li2019}.}

For the Bi$_4$I$_4$, density functional theory (DFT) and GW calculations predicted that $\beta$-phase ($T > T_P$)  hosts electronic bands structure with non-trivial topology \cite{liu2016weak,autes_novel_2016}. Such predictions are so far supported by Angle-Resolved Photoemission Spectroscopy (ARPES)~\cite{autes_novel_2016, noguchi_weak_2019,huang2021room}, reporting the observation of Dirac band within the bulk band gap only for the $\beta$ phase. However, the classification of the emergent topological phases in Bi$_4$I$_4$ crystal structures is not conclusive in several theoretical predictions. 
In the case of the $\beta$ phase, a consensus on its topological classification has not been found and both strong \cite{autes_novel_2016} or, the least studied, weak topological insulating state have been proposed~\cite{noguchi_weak_2019,huang2021room}. For the $\alpha$-Bi$_4$I$_4$ ($T < T_P$), several works pointed to this phase as a trivial topological insulator \cite{autes_novel_2016,noguchi_weak_2019}, whereas others put forward the possibility of $\alpha$-Bi$_4$I$_4$ being a rare higher-order topological insulator hosting a hinge state \cite{yoon_quasi-one-dimensional_2020, huang2021room}, the latter being considered as a possible material to mediate topological superconductivity \cite{liu2022gate}.

Since the structural phase transition between the two phases ($\alpha$ and $\beta$) occurs close $\sim T_P$, %($T_P \approx 300$~K),
any experiment carried out at room temperature to investigate physical properties and relate to specific signatures of distinct topology will present the challenge to guarantee which is the actual investigated phase. To overcome this issue, what is often reported to stabilize the $\beta$-phase and to perform comparative investigations between the $\beta$ and $\alpha$-Bi$_4$I$_4$, is to freeze the crystalline structure by quenching the crystals in cold water \cite{von1978bismuth} or liquid nitrogen \cite{noguchi_weak_2019} right after the growth. By doing that, the two phases ($\alpha$ and a metastable $\beta$) were assumed to become stable for $T<T_P$. 
This procedure is not adopted, or at least not explicitly stated, in all the reports that studied the $\beta$-phase \cite{autes_novel_2016,huang2021room, pisoni_pressure_2017,wang2021quantum}.

Considering the martensitic nature of the phase transition between the $\alpha$ and $\beta$ phases, it is expected that quenching and ageing of the samples could influence the phase transformation~\cite{kustov2015effect,romero1997quenched}. Previous works addressed methodologies to improve the growth process to obtain bigger samples with a single phase \cite{autes_novel_2016,noguchi_weak_2019}, but a study of the phase stability of Bi$_4$I$_4$ 
has not been reported so far. 
In this work, we carried out a systematic structural investigation of Bi$_4$I$_4$ across a series of controlled thermal treatments starting from $\alpha$ - Bi$_4$I$_4$ single crystals annealed at $T>T_P$, and then quenched. We found that Bi$_4$I$_4$ does not consistently reach a metastable $\beta$-phase upon quenching,  displaying a persistent $\alpha$-phase or quickly returning to it in a few hours. The site occupancy number (SON) obtained from the analysis of the XRD data before and after the thermal cycle indicates that the observed instability is concomitant with an increase in native defects. The latter scenario is also consistent with the observation of a change in the temperature dependence of electrical resistivity before and after the quenching process. Our DFT calculations support this scenario, finding that bismuth antisites are the predominant native defect in both $\alpha$ and $\beta$ phases under Bi-rich conditions, followed by vacancies, iodine antisites, and bismuth interstitials. We argue that the presence of these defects plays a key role in the phase transition of Bi$_4$I$_4$ due to its martensitic nature.

%%%%%%%%%%%%%%%%%%%%%%%%%%%%%%%%%%%%%%%%%%%%%%%%%%%%%%%%%%%%
%%%%%%%%%%%%%%%%%%%%%     METHODS      %%%%%%%%%%%%%%%%%%%%%
%%%%%%%%%%%%%%%%%%%%%%%%%%%%%%%%%%%%%%%%%%%%%%%%%%%%%%%%%%%%
%%%%%%%%%%%%%%%%%%%%%%%%%%%%%%
%%%%%%     FIGURE 1     %%%%%%
%%%%%%%%%%%%%%%%%%%%%%%%%%%%%%

\begin{figure*}[!ht]
    \includegraphics[width = 17cm]{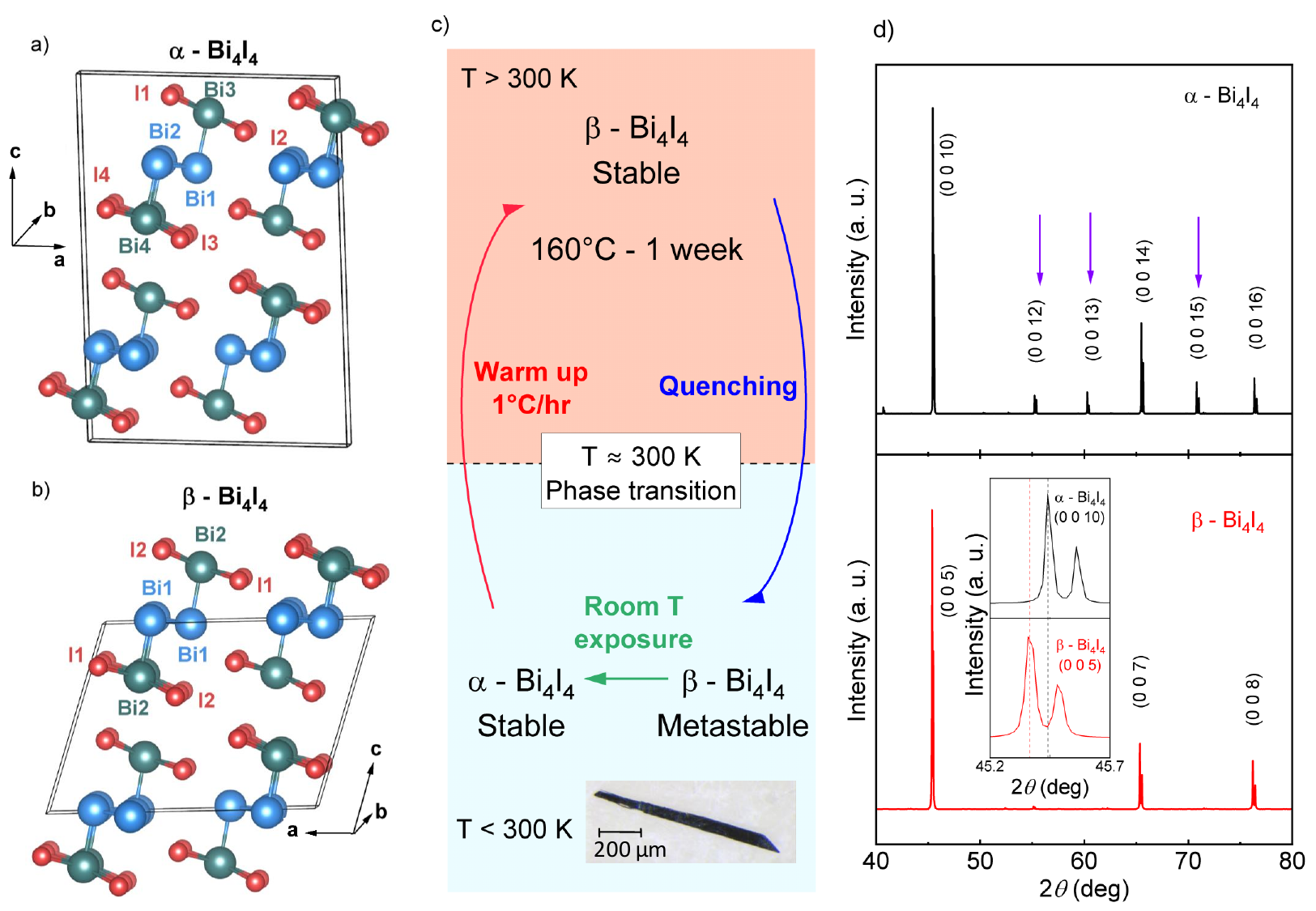}
    \caption{Unit cells of the (a) $\alpha$- Bi$_4$I$_4$ and (b) $\beta$ - Bi$_4$I$_4$. We denote the nonequivalent crystallographic sites as Bi$n$ and I$n$ for both phases. In the case of $\alpha$, $n=1,2,3,4$; whereas in the case of $\beta$, $n=1,2$. Different colours distinguish the inner (blue) and outer (green) bismuth atoms. (c) Scheme of the expected phase evolution of Bi$_4$I$_4$ during the thermal treatment procedure. The inset shows a photograph of a representative as-cast sample. (d) XRD profile simulations of both phases, where purple arrows indicate characteristic peaks of the $\alpha$-phase. Inset illustrates the shift of the $\beta$ - Bi$_4$I$_4$ peak, where the primary and secondary reflections can be distinguished in detail.}
    \label{fig:cells}
\end{figure*}

\section{METHODS} 

Bi$_4$I$_4$ single crystals were obtained by Chemical Vapour Transport (CVT) technique, as described in previous works \cite{filatova2007electronic}. A powder mixture of HgI$_2$ and pure bismuth with a molar ratio of 1:2 was used as starting materials. The mixture was then sealed on a quartz tube and then placed slightly tilted in a two-zone furnace with a thermal gradient of 280 / 210$^\circ$C, with the starting materials at the hot and lower sides. After one week, the crystals were cooled down slowly to finally obtain $\alpha$-Bi$_4$I$_4$ (Fig.\ref{fig:cells}(a)) single crystals.
As cast $\alpha$ - Bi$_4$I$_4$ single crystals were analysed by energy-dispersive x-ray spectroscopy (EDS) using a Tescan Mira3 Field Emission Scanning Microscope equipped with an ENAX Octane Elite system. Single crystals of $\alpha$-Bi$_4$I$_4$ were obtained with average size 1.5 $\times$ 0.2 $\times$ 0.003 mm$^3$ and atomic ratio  Bi:I 51\%:49\%, i.e. rich in bismuth, similar to the values reported in previous works \cite{pisoni_pressure_2017,wang2021quantum}. 

For the thermal treatment, $\alpha$ - Bi$_4$I$_4$ samples were selected from the same batch and sealed in a glass tube with an argon atmosphere just below atmospheric pressure to avoid surface contamination and ensure thermal contact with the outer environment. The sealed samples were placed into a muffle furnace, heated from 50$^\circ$C to  160$^\circ$C using a ramp of 1$^\circ$C/min, and maintained at 160$^\circ$C for one week, reproducing the time at which the crystals are kept during the process of growth. Then, samples were quenched through the rapid cooling of the glass tube using iced water, aiming at obtaining the $\beta$ - phase (Fig. \ref{fig:cells} (b)). The thermal treatment is schematically shown in Fig.\ref{fig:cells} (c). Unless explicitly stated, the resulting crystals have been maintained below room temperature (T $\approx 5^\circ$C) after quenching until beginning the XRD measurements and handled inside a cooler not to overcome $T_P$ while transporting them. 

Measurements were obtained by x-ray diffraction (XRD) at room temperature ($\sim 25^\circ$C), before and after thermal treatment, using a Bruker D8 X-ray diffractometer with Cu $K\alpha$ radiation ($\lambda_1 = 1.5406$ $\text{\AA} $, $\lambda_2 = 1.5444$ $\text{\AA}$ and $I_2/I_1 = 0.5$), in a Bragg-Brentano geometry, which also confirmed the natural cleavage orientation of our crystals. To optimize the statistics and reduce the time to obtain each diffraction pattern, we rotated our samples at 30 rpm with a step of $\Delta 2\theta = 0.02^\circ$ and a time of measurement per step of 2 seconds. Each measurement took around 1 hour to be completed, which is the minimum time after thermal treatment we achieved. The analysis of the XRD profiles was performed using the Rietveld Refinement method as implemented in the FullProf software \cite{rodriguez1990fullprof}. Electrical resistivity was measured within 8 - 295 K by the four-probe method in selected $\alpha$-Bi$_4$I$_4$ crystals. 25 $\mu$m gold wires were placed directly onto the sample using silver paint. Measurements were performed in an Advanced Research System DE-202N cryocooler.

%%%%%%%%%%%%%%%%%%%%%%%%%%%%%%
%%%%%%     FIGURE 2     %%%%%%
%%%%%%%%%%%%%%%%%%%%%%%%%%%%%%
\begin{figure*}[!ht]
    \centering
    \includegraphics[width = 12.9cm]{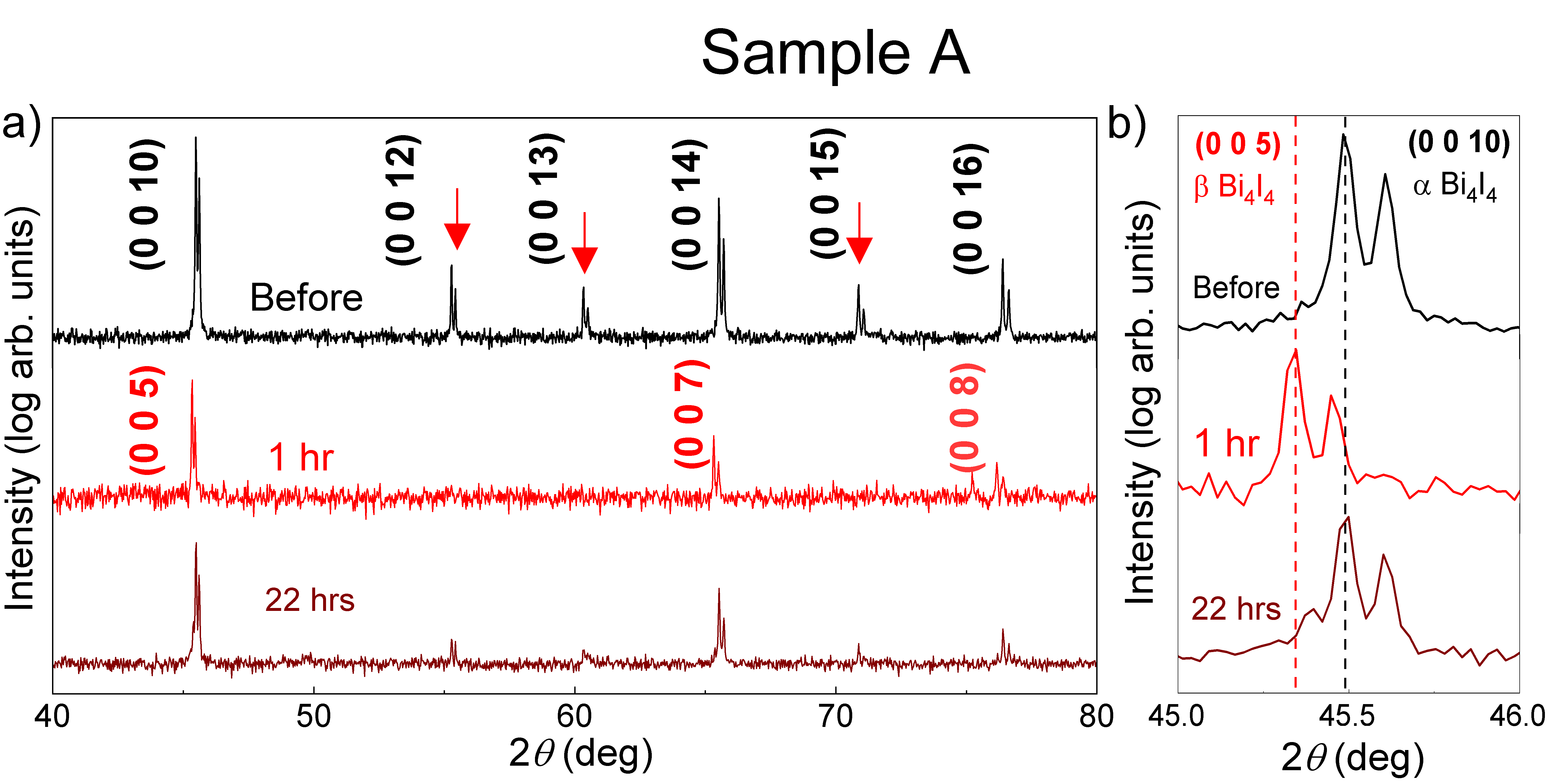}
    \includegraphics[width = 12.9cm]{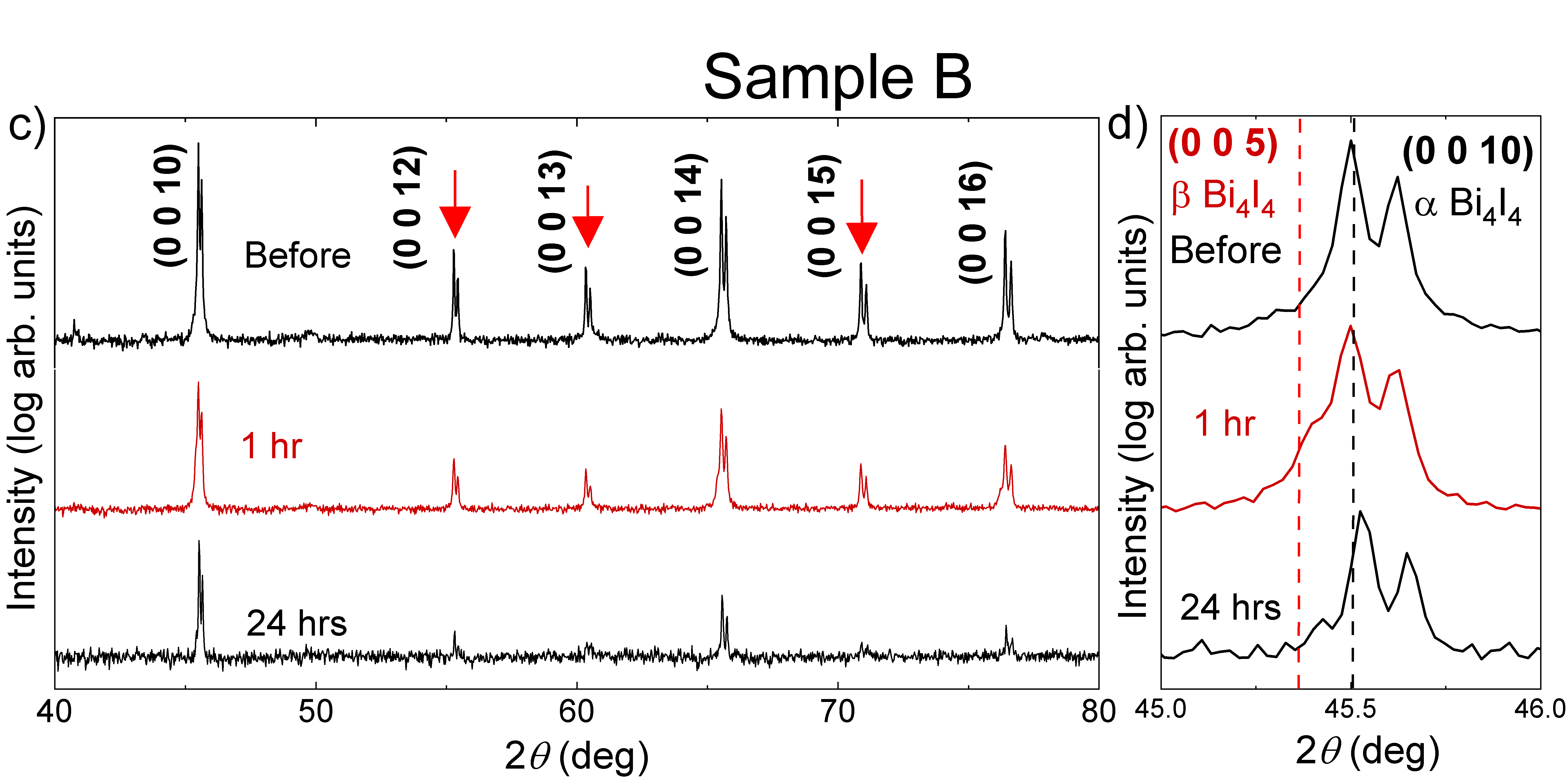}
    \caption{(a) XRD profiles before and after thermal treatment of sample \textbf{A}. Red downward arrows point to the characteristic reflections identifying the $\alpha$ phase. Those peaks allow discerning between the two phases. (b) shows a close-up to the peak around 45$^\circ$, which corresponds to (0 0 10) and (0 0 5) peaks of the $\alpha$ and $\beta$ phases, respectively.   (b)  (c) and (d) same as (a) and (b) but for the sample \textbf{B}.}
    \label{fig:figure2}
\end{figure*}

From the theoretical side, DFT calculations were performed within the Perdew-Burke-Ernzehof generalized gradient approximation (GGA-PBE)~\cite{pbe-gga} including van der Waals corrections (vdW-DF)~\cite{vdw} and spin-orbit coupling (SOC), as implemented in the Vienna \textit{Ab initio} Package (VASP)~\cite{vasp1,vasp2}. We employed projector augmented wave (PAW)~\cite{paw} potentials and relaxed the structures until the forces on each atom were $<0.01$ eV/$\text{\AA}$, and total energies are converged within a $1\times 10^{-6}$ eV criterion. For the plane-wave basis we used an energy cutoff of 400 eV. The total energies were calculated using a \textit{k}-point mesh of $4 \times 3 \times 2$.

%%%%%%%%%%%%%%%%%%%%%%%%%%%%%%%%%%%%%%%%%%%%%%%%%%%%%%%%%%%%
%%%%%%%%%%%%%%%%%%%%%     RESULTS      %%%%%%%%%%%%%%%%%%%%%
%%%%%%%%%%%%%%%%%%%%%%%%%%%%%%%%%%%%%%%%%%%%%%%%%%%%%%%%%%%%

\section{RESULTS}

\subsection{$\beta$-Bi$_4$I$_4$ phase instability}

The XRD analysis leads to the determination of the lattice parameters $a = 14.24(2)$~$\text{\AA}$, $b=4.430(4)$~$\text{\AA}$, $c=19.9682(4)$~$\text{\AA}$ and $\beta = 92.98^\circ$. As can be seen in Figs. \ref{fig:cells} (a) and (b), the $\alpha$ and $\beta$ phases share the same C2/m space group, but they have distinct $c$ parameters, namely $c_\alpha$ ($\alpha$-phase) and $c_\beta$ ($\beta$-phase), whose ratio is $c_{\alpha}/c_{\beta} \approx 1.9$. As $c_\alpha > c_{\beta}$, some plane reflections can be observed only on the $\alpha$ phase. The distinction between the two phases becomes then possible through the comparative analysis of the diffraction spectra. It is instructive to look at a simulation of the XRD for the ideal and pristine phases in the 2$\theta$ range that is explored in our experiments [Fig.~\ref{fig:cells} (d)]: the $\alpha$ phase is expected to display characteristic diffraction peaks (downward arrows in the Figure) that are not present for the $\beta$ phase. Additionally, the inset of Fig.~\ref{fig:cells} (d) (lower panel) shows a zoom into the peak at around 45.5$^\circ(2)$, corresponding to the (0 0 10) peak of the $\alpha$ and the (0 0 5) of $\beta$ phase; there, due to the mismatch in the value of $c$, a shift to the left can be resolved for the $\beta$-phase. The position of the diffraction peaks and the evolution of the double-peak structure at 45.5$^\circ$ will be used to analyse our experimental diffractograms. 
%The simulation considered the primary and secondary radiation sources to be compared with the experimental results. 

It is worth mentioning that peaks for the (0 0 5) and (0 0 7) planes of the $\alpha$-phase, placed at $2\theta \approx 22.3^\circ$ and $31.5^\circ$ [not shown in the simulation in Fig. \ref{fig:cells}(d)], are typically used in the literature to distinguish between the $\alpha$ and $\beta$ structures~\cite{huang2021room,noguchi_weak_2019,pisoni_pressure_2017,wang2021quantum}. However, a careful XRD analysis of our samples showed that those characteristic diffraction peaks appear even when a mixture of phases is present and that the two contributions might not be resolved. As mentioned above, because $c_\alpha$ is not exactly double $c_\beta$, $\beta$ phase (0 0 $l$) peaks are expected to appear slightly shifted to the left concerning the ones corresponding to the (0 0 $2l$)  $\alpha$ phase peaks, but the effect becomes more pronounced at higher angles. 
For these reasons, to make the contributions of the two neighbours $\alpha$ and $\beta$ peaks identifiable, we chose to investigate the $40$ to $80^\circ$ $2\theta$ window.

A series of Bi$_4$I$_4$ samples from the same batch was analysed by XRD before and after thermal treatment as described in the Methods section. Four selected samples (labelled \textbf{A}, \textbf{B}, \textbf{C}, \textbf{D}), representative of the findings, are reported and discussed in this work. All specimens were in the $\alpha$ phase before the thermal treatment (see supplemental material \cite{SM}). Overall we find that the annealing and successive quenching did not have a consistent effect in stabilizing the $\beta$ phase, resulting, in the majority of attempts, in a combination of phases. 

Figure \ref{fig:figure2} shows the time evolution of the XRD profile for sample \textbf{A}. The XRD of the $\alpha$ phase before the thermal treatment is depicted in Fig. \ref{fig:figure2}(a), where red downward arrows point to the diffraction peaks for this phase. Right after quenching [1hr, see the red curve in panel (a)], the $\beta$ phase was obtained at $T < T_P$: the (0 0 12), (0 0 13), and (0 0 15) peaks are not detected within the resolution. The refined lattice parameters of the obtained $\beta$ - Bi$_4$I$_4$ single crystal are $a = 14.37(1)$~$\text{\AA}$, $b=4.13(3)$~$\text{\AA}$, $c=10.487(9)$~$\text{\AA}$ and $\beta= 107.7^\circ$. After room temperature exposure ($\sim 25^\circ$C), the percentage of the $\beta$-phase reduced drastically: the characteristic peaks of the $\alpha$ phase become dominant just after 22 hrs. The latter becomes even clearer by analyzing the shape evolution of the 45.5$^\circ$ peak, from which we estimate a remaining percentage of approximately 10\% of the $\beta$ phase after 22 hrs. Interestingly, after repeating the same thermal treatment procedure on the same sample (\textbf{A}), we did not detect any contribution of the $\beta$ - phase at $T < T_P$ within the XRD resolution.

The time evolution of the XRD profile of sample \textbf{B} is reported in Fig.~\ref{fig:figure2}(c) and (d). 
Comparing the diffractograms before the thermal treatment and right after quenching (1 hr), we can observe that, unlike in the case of sample \textbf{A}, the $\alpha$ phase peaks are present already after 1 hr. To detect the partial presence of the $\beta$-phase, we had to inspect %his can be understood as a quick phase transition or a blocking of the $\beta$ phase stability after thermal treatment.
%Inspecting 
the evolution of the 45.5$^\circ$ peak [Fig.~\ref{fig:figure2}(d)]. There, a minor contribution of the $\beta$ phase can still be identified after (1 hr), where the peak displays a shoulder associated with this phase, whose contribution can be estimated to be about 20\%. After 24 hrs, a $\beta$ contribution is not detectable. Our findings point to a very quick relaxation of the $\beta$-phase towards the $\alpha$-phase even after a time as short as one hour.

Finally, samples \textbf{C} and \textbf{D} (see Sec. I in Supplemental Material \cite{SM}) persisted in the $\alpha$-phase after thermal treatment and quenching. For both these samples, a careful inspection of the reflection peak around $45.5^\circ$ did not reveal a detectable contribution of the $\beta$-phase.

\begin{figure}[!h]
    \centering
    \includegraphics[width=8.6cm]{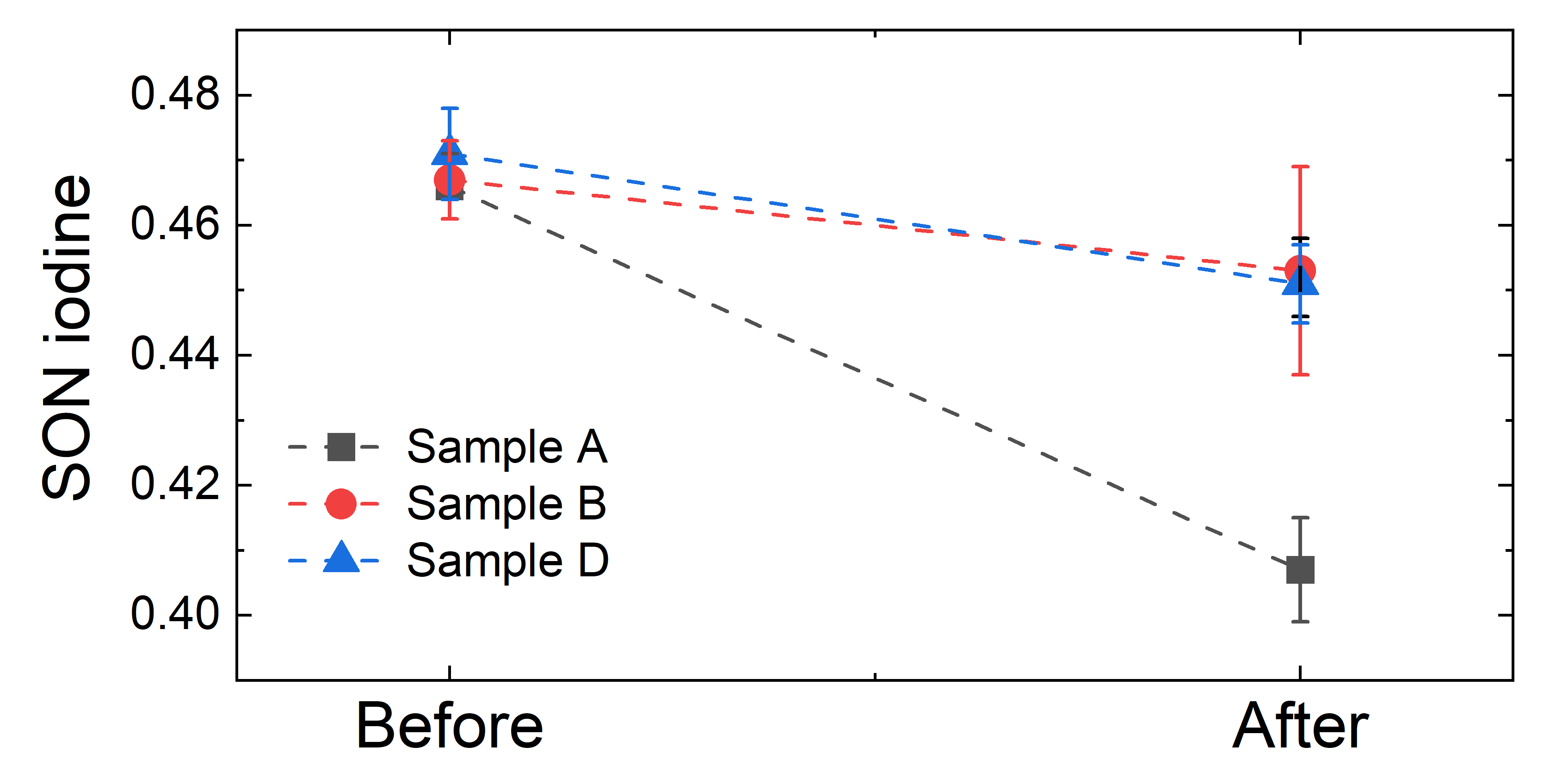}
    \caption{Site occupancy number (SON) of iodine sites for samples \textbf{A}, \textbf{B} and \textbf{D} before and after thermal treatment. A reduction to values lower than the ideal SON = 0.5 corresponds to the formation of vacancy defects. } 
    \label{fig:SON}
\end{figure}

Eventually, all investigated samples returned to a $\alpha$ phase after a long enough exposure at room temperature. We then attempted to include point defects in the Rietveld analysis following the procedure taken in \cite{kaufmann2023investigation_FeGa3}. To tentatively estimate the evolution of defects upon thermal treatment, we refined the $\alpha$ phase XRD profiles of samples \textbf{A}, \textbf{B}, \textbf{C}, and \textbf{D} before (as-cast) and after the heating/quenching process. In the case of samples \textbf{A} and \textbf{B} for the post-treatment data set, we selected the XRD profile after they entirely transitioned back to the $\alpha$ phase.
Our results were not conclusive for interstitial and antisite defects, whose selective inclusion did not improve the refinements within the resolution. 
On the other hand, in the as-cast crystals, we found a sizable presence of iodine vacancies, which increase after treatment in samples \textbf{A}, \textbf{B} and \textbf{D}, as shown in Fig.~\ref{fig:SON}.
This points to a relevant role of the thermal trajectory of the samples in the generation of defects, especially iodine vacancies, both before and after treatment.  Table~\ref{tab:R_Bragg} displays improvement of the R$_{Bragg}$ factors, which depend only on the structural parameters \cite{toby2006r}, obtained in the refinements. In this case, a formula Bi$_4$I$_{4-x}$ can be used, where $x$ can be obtained from the SON values (Fig.~\ref{fig:SON}) by adding the refined SON for each atomic species and considering that the chemical proportion in the sample is given by that sum. The value of $x$ reaches a maximum value of 0.09 for sample \textbf{A} after treatment. 

It is important to remark that our analysis does not exclude the possibility of having interstitial or antisite defects as well as the formation of more than one defect at the same time, but higher-resolution XRD experiments are needed to explore the presence of these defects further quantitatively \cite{yin2018extending_XRD_defects}.

\begin{table}[!h]
    \centering
       \caption{Resulting $R_{Bragg}$ structure factors of refinements after thermal treatment considering iodine vacancies. }
    \begin{ruledtabular}
    \begin{tabular}{c c c c }
        & Sample A & Sample B & Sample D \\ \hline
        Pristine & 16.8 & 28.9  & 16.4 \\
        I Vacancy & 16.7 & 28.8  &  16.4\\
    \end{tabular}
    \end{ruledtabular}
    \label{tab:R_Bragg}   
\end{table}
%%%%%%%%%%%%%%%%%%%%%%%%%%%%%%
%%%%%%     FIGURE 3     %%%%%%
%%%%%%%%%%%%%%%%%%%%%%%%%%%%%%

\subsection{Electrical resistivity in Bi$_4$I$_4$}
\begin{figure*}[tp]    
\includegraphics[width = 17cm]{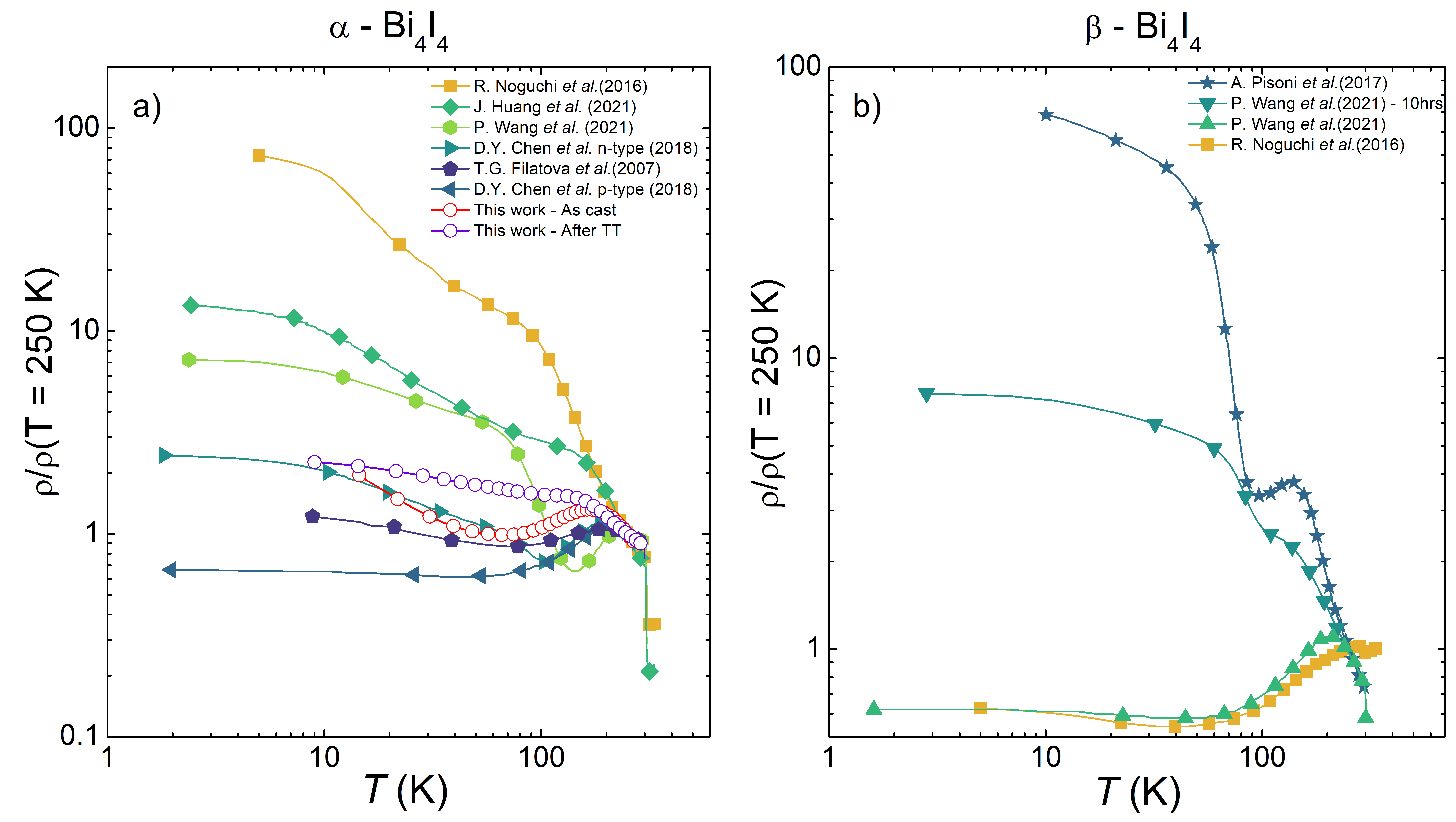}
    \caption{Comparison of electrical resistivity measurements for the (a) $\alpha$-Bi$_4$I$_4$ and (b)$\beta$-Bi$_4$I$_4$, respectively. The open circles refer to data of this work for the $\alpha$ phase, as cast and after thermal treatment (TT) data was obtained from~\cite{noguchi_weak_2019,wang2021quantum,filatova2007electronic,chen2018quantum,pisoni_pressure_2017} and the lines are just guides for the eyes. }
    \label{fig:FigRho}
\end{figure*}

Electrical resistivity measurements down to low temperatures (1.5$-$300~K) can be found in the literature on both the $\alpha$ and the (quenched) metastable $\beta$ - Bi$_4$I$_4$ phases and are depicted in Figs. \ref{fig:FigRho}(a) and (b), respectively. Our measurements for the $\alpha$-Bi$_4$I$_4$ before and after thermal treatment are also included in Fig. \ref{fig:FigRho}. We did not measure $\rho(T)$ in the metastable $\beta$ phase at low temperatures, as all investigated samples transitioned back to the $\alpha$ phase.

According to the previous works cited in  Fig. \ref{fig:FigRho}, the $\beta$ phase was stabilized at $T<T_P$ and, thus, it was possible to measure its electric transport at low temperatures.
 Interestingly, $\rho(T)$ shows a large variation of behaviour in both phases, displaying a saturation at values ranging of 2 orders of magnitude at the lowest temperatures to the value at 300~K, pointing to an indisputable sample dependence.  
In particular, let us observe $\rho(T)$ data for both phases from Noguchi and collaborators \cite{noguchi_weak_2019} displayed in Figs. \ref{fig:FigRho} (a) and (b) (in dark yellow), respectively. In that case, the low-temperature divergence of the $\alpha$-phase and the saturation at lower values for the $\beta$ phase might be suggestive of a possible signature of a distinct topological classification of the electronics bands \cite{noguchi_weak_2019} given by the structural phase. Nevertheless, the general comparison of the available data does not support a consensus about the expected fingerprint of the two phases in electric transport. 

 Finally, our $\rho(T)$ measurement for the as-cast sample in the $\alpha$ phase, selected from the same initial batch and non-thermally treated, is in agreement with the values reported by Chen \emph{et al.} \cite{chen2018quantum} and Filatova \emph{et al.} \cite{filatova2007electronic} [in figure \ref{fig:FigRho} (a)].  However, $\rho(T)$ in the same $\alpha$ phase sample after quenching shows a small but sizable variation in the temperature dependence with a less pronounced upturn at low temperatures, pointing to the effect of the thermal treatment on the crystal. Our experimental setup allows the temperature control up to 295~K, therefore, the expected jump in the electrical resistivity at the phase transition is not observed for our crystals.
 %\textcolor{blue}{It is worth to point that we measured $\rho(T)$ up to 295K to avoid the phase transition to the $\beta$-Bi$_4$I$_4$ and keep our samples in the $\alpha$ phase.}

Generally, native defects have important effects on the electronic properties of topological materials. For instance, defects can induce unintentional doping and give rise to scattering centres, affecting these materials' electrical resistivity. In the case of our Bi$_4$I$_4$ samples, we have an indication of an increased number of iodine vacancies after thermal treatment, which is compatible with the variation in the temperature dependence in $\rho(T)$ in the same specimens.

\subsection{Native defects within DFT calculations}

To support our conclusion on the formation of the native defects accompanying the phase instability and change in $\rho(T)$ after thermal treatment, we studied their relative abundance in Bi$_4$I$_4$ employing first-principles calculations.

\begin{figure}[!h]
    \centering
    \includegraphics[width=8.6cm]{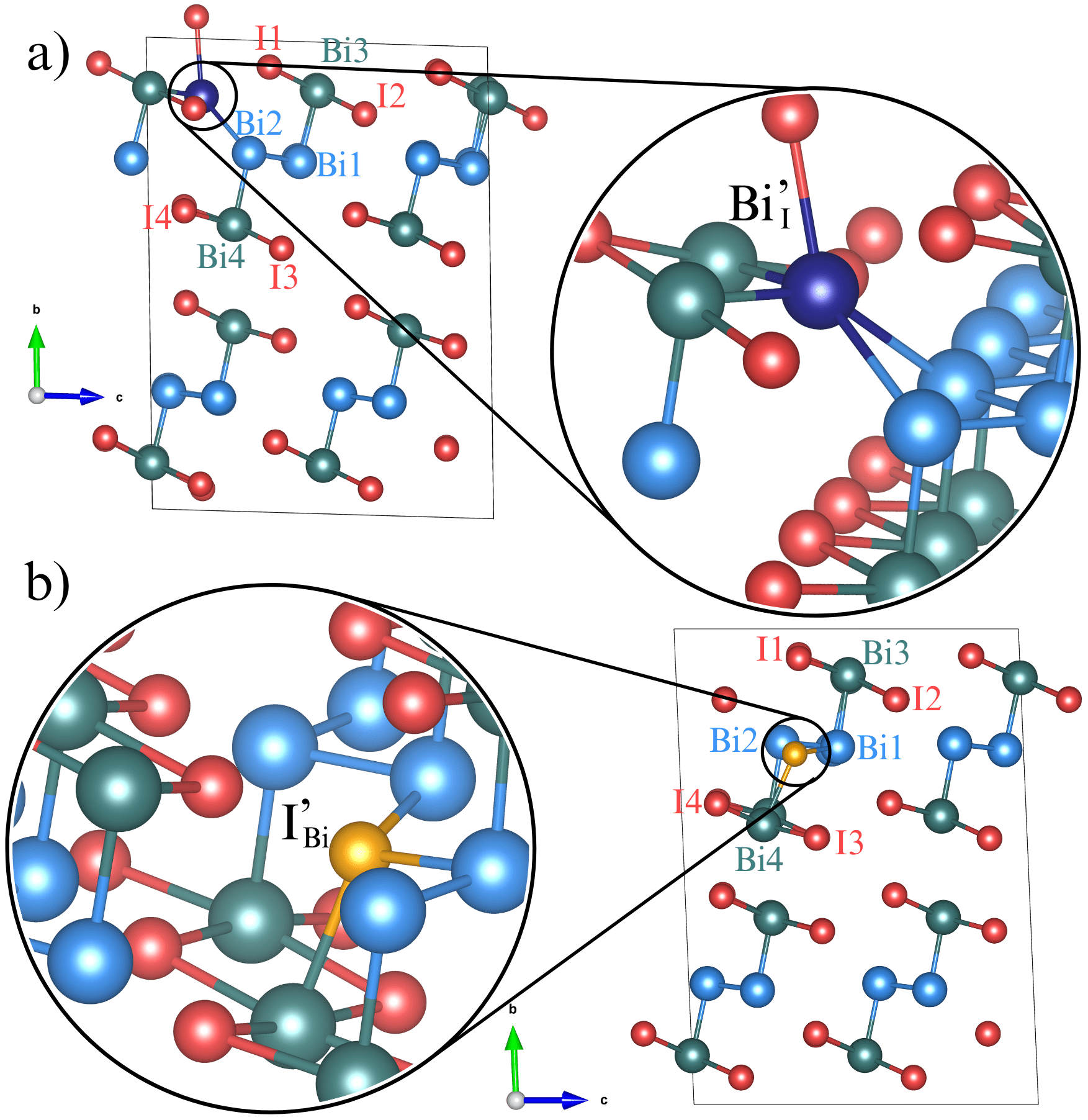}
    \caption{Relaxed structure of $\alpha$-Bi$_4$I$_4$ supercell with (a) Bismuth and (b) Iodine antisites. The primitive cell of $\alpha$-Bi$_4$I$_4$ used to generate the supercell is shown in supplementary material \cite{SM}. Dark blue and yellow spheres represent defects.}  
     \label{fig:figure4Struca}
\end{figure}

\begin{figure*}[!htb]
    \centering
      \includegraphics[scale=0.075]{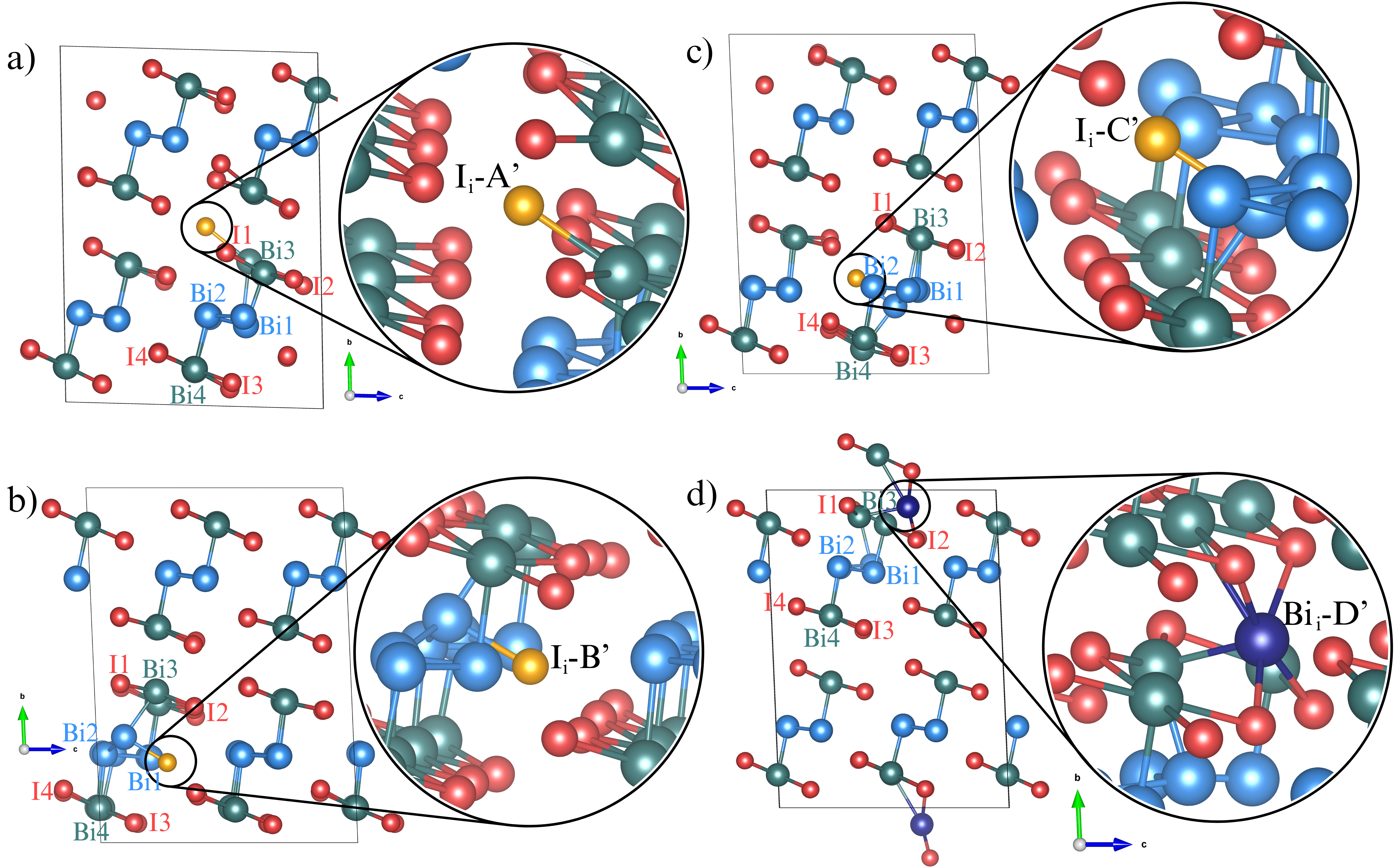}
        \caption{Relaxed structures of $\alpha$-Bi$_4$I$_4$ supercell with interstitial defects. In (a),(b) and (c) we present iodine interstitial configurations labelled as I$_i$-A', I$_i$-B' and I$_i$-C', respectively. In (d), we display the relaxed structure of the most stable interstitial bismuth, Bi$_i$-D'. Dark blue and yellow spheres represent the defects.}
        \label{fig:figure4Strucb}
\end{figure*}

\begin{figure}[!htb]
    \centering
      \includegraphics[scale=0.065]{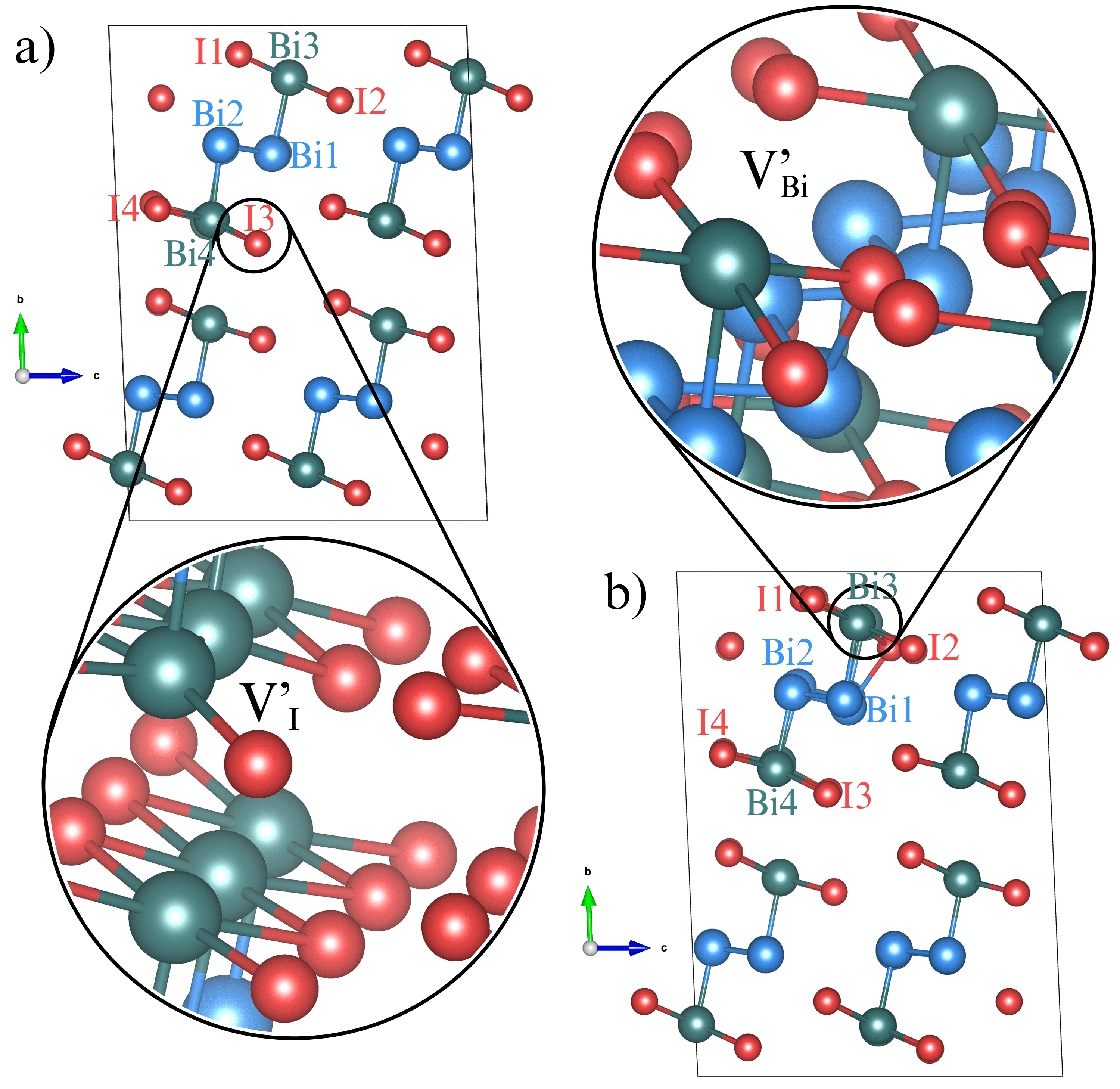}
        \caption{Relaxed structures of $\alpha$-Bi$_4$I$_4$ supercell with (a) iodine and (b) bismuth vacancies.}
        \label{fig:figure4Strucc}
\end{figure}

In particular, we calculated the formation energies of antisites, interstitials, and vacancies in both $\alpha$ and $\beta$ phases, under Bi-rich conditions, consistent with our experimental conditions. We limit our theoretical analysis to the case of neutral defects, which have formation energies given by~\cite{van2004first}
\begin{equation} 
E_{D}^{f} = E_{tot}^{D}-E_{tot}^{bulk} - \sum_{i}n_{i}\mu_{i},
\end{equation}
where $E_{tot}^{D}$ and $E_{tot}^{bulk}$ are the DFT obtained total energies of the defective and pristine configurations for the same supercell, respectively. $n_{i}$ refers to the number of exchanged atoms with reservoirs characterized by chemical potentials $\mu_{i}$, where the sub-index $i$ refers to the atomic species: Bi or I. Our choice of $\mu_{Bi}$ was to obtain it from the bulk bismuth, which exhibits $P2_1/m$ (No. 11) symmetry. Therefore, from the thermodynamic equilibrium condition, $\mu_{I} = (\mu_{Bi_{4}I_{4}} - 4\mu_{Bi})/4$. $\mu_{Bi_{4}I_{4}}$ is the calculated total energy per formula unit.  The obtained formation energies of the most stable configurations, regarding the same type of defects, are displayed in Table~\ref{tab:form_ene}. The corresponding relaxed structures of the supercells with antisites, interstitials, and vacancies in the $\alpha$ phase are shown in Figs.~\ref{fig:figure4Struca}, ~\ref{fig:figure4Strucb}, and ~\ref{fig:figure4Strucc}, respectively. The relaxed structures of the defects in the $\beta$ phase are shown in Figs. S4, S5 and S6 of the supplemental material.

\begin{table*}[t]
    \centering
       \caption{Calculated formation energies in Bi-rich conditions of neutral antisite, interstitials, and vacancies in the $\beta$ and $\alpha$ phases of Bi$_4$I$_4$. The antisites are denoted as Bi$_I$ and I$_{Bi}$ followed by the site index (between parentheses). Interstitials, in turn, are denoted by Bi$_{i}$ and I$_{i}$, followed by a letter that indicates interstitial positions as in Fig. \ref{fig:figure4Strucc} for the $\alpha$ phase (for the $\beta$ phase see Supplemental material \cite{SM}). Vacancies are denoted by V$_{I(Bi)}$ followed by the site index corresponding to the removal of one atom (between parentheses).}
    \begin{ruledtabular}
    \begin{tabular}{c c c c c}
       Defect type & Configuration ($\beta$-phase) & $E^{f}_{D}$ ($\beta$-phase) (eV) & Configuration ($\alpha$-phase) & $E^{f}_{D}$ ($\alpha$-phase) (eV)\\ \hline
       \multirow{2}{*}{Antisites} & Bi$_I$ (\textit{I1}) & 0.53 & Bi'$_I$  (\textit{I2})  & 0.41 \\ 
       & I$_{Bi}$ (\textit{Bi1}) & 1.18 & I'$_{Bi}$ (\textit{Bi2}) & 1.08  \\ \hline
       \multirow{5}{*}{Interstitials} & I$_i$-A & 1.59 & I$_i$-A' & 1.67\\
        & I$_i$-B & 1.57 & I$_i$-B' & 1.80  \\
        & I$_i$-C & 1.56 & I$_i$-C'&  1.72\\
        & I$_i$-D & 1.72 & Bi$_i$-D'& 1.07 \\
        & Bi$_i$-E & 1.19 & &  \\ \hline
        \multirow{2}{*}{Vacancies} & V$_I$ (\textit{I2}) & 1.02 & V'$_I$ (\textit{I3})  & 0.96\\
        & V$_{Bi}$(\textit{Bi2}) & 0.94  & V'$_{Bi}$(\textit{Bi3}) & 0.93 \\     \end{tabular}
    \end{ruledtabular}
    \label{tab:form_ene}   
\end{table*}

According to the calculated formation energies, in Bi-rich conditions, the %predominant
most stable defect in Bi$_4$I$_4$ is the bismuth antisite (Bi$_I$), followed by bismuth(iodine) vacancies (V$_{Bi(I)}$). The former has a slightly smaller formation energy, $\Delta E_{D}^{f} \approx 0.1$~eV, in the $\alpha$ phase compared to the same defect in the $\beta$ phase. The vacancies, in turn, have formation energies of around 1~eV in the $\alpha$ and $\beta$ phases. 
We observe that in the case of bismuth antisite, the energetic preference of Bi is to replace the iodine at site \textit{I2} (Fig.~\ref{fig:figure4Struca}(a)). 
As can also be noticed in Fig.~\ref{fig:figure4Strucc}, the iodine vacancy does not significantly alter the local structure, while the bismuth vacancy leads to a sizable local distortion. 

It is also important to mention that our calculations indicate that Bi atoms are slightly less stable in \textit{I4} sites, with the former being $0.03$~eV more stable than the latter. 

The iodine (I$_{Bi}$) antisite and Bi-interstitials exhibit similar formation energies and are expected in lower concentrations in Bi-rich conditions. As can be noticed in Table~\ref{tab:form_ene}, iodine antisites have formation energies of around 0.7~eV larger than the Bi$_I$ defects, indicating that I$_{Bi}$ are unlikely to be found in our growth conditions. The interstitial bismuths are around 0.66~eV less stable than Bi antisites in both $\beta$ and $\alpha$ phases, 
though they are more stable than iodine interstitials.

Finally, we mention that among the interstitial iodine, our calculations indicate that the configurations I$_i$-A' and I$_i$-C' (Figs.~\ref{fig:figure4Strucb}(a) and (c)) are the energetically more stable.
It is important to mention that these defects can give rise to grain boundary pinning and chemical disordering, which can have important effects on the martensitic transition of Bi$_4$I$_4$ \cite{yang2022role}.

\section{DISCUSSION}
One of the main findings of this work is that the metastable $\beta$-Bi$_4$I$_4$ phase (Fig.~\ref{fig:figure2}) returns to the $\alpha$-Bi$_4$I$_4$ relatively quickly when the quenched crystals are exposed to room temperature, raising a challenge on how to maintain and guarantee the expected phase during experiments requiring post-growth preparation and handling of the specimens. 

Wang \emph{et al.} \cite{wang2021quantum} (cf. Fig. \ref{fig:FigRho}) investigated electrical transport in $\beta$-Bi$_4$I$_4$ and also observed that $\rho(T)$ changes over time, as a consequence of room temperature exposure: freshly investigated samples showed saturation at low temperature of the electrical resistivity, while after 10 hours exposure the curve diverges resembling the expected behaviour for the $\alpha$ phase. They attribute the effect to possible chemical instability for exposure to moisture. Our results suggest an interpretation based on intrinsic instability of the $\beta$-Bi$_4$I$_4$ for $T<T_P$, related to the formation of native defects along the thermal history of the specimen. The appearance of bismuth interstitial defects was found in Bi$_2$Se$_3$ \cite{tumelero2016role,urazhdin2004surface} and Bi$_2$Te$_3$ \cite{callaert2019interstitial}, playing an important role on the emergence of specific signatures of topological features in physical properties. 
Antisite defects were recently suggested to have a relevant role in the formation of impurities bands in narrow-gap semi-conductors \cite{kaufmann2023investigation_FeGa3}. In the case of Bi$_4$I$_4$, native defects can also be relevant in the determination of the charge carriers at the Fermi level. The variation of $\rho(T)$ in the $\alpha$ phase after quenching with respect to the as-cast samples suggests that native defects modify the transport properties of Bi$_4$I$_4$.

Our findings are also in agreement with the observations done by D. Mu \emph{et al.} \cite{mu2023role}, who reported the presence of iodine vacancies and the coexistence of the $\alpha$ and $\beta$ - phases in Bi$_4$I$_4$ single crystals,
%, by Scanning Tunneling Microscopy (STM), 
the latter related to the presence of hollow defects at the nanoscale. According to their work, hollow defects allow the formation and stability of the $\beta$ phase, while the hollow density varies between samples of the same batch. 
%This hypothesis can explain the sample dependence of our results.  

It is important to observe that even in as-cast $\alpha$-Bi$_4$I$_4$ samples, we could quantify a sizable presence of iodine vacancies, which are among the most stable defects likely to be formed according to our DFT calculations together with Bi-antisites and interstitials. %The latter two could not be identified in the refinements of the XRD data, but we can't exclude that higher resolution investigation can detect their presence and evolution. 
Therefore, we infer that even the initial growth conditions can determine an initial level of native defects and that they increase through successive trajectories across the martensitic transition. These points defects can affect the associated martensitic phase transition \cite{roitburd1979nature,hornbogen1985effect} or even block the transformation~\cite{hennig2005impurities}, when the crystals undergo a quenching process.
\begin{table*}[t!]
  \caption{Summary of the growing parameters used in this work and by different authors in the literature.}
  \begin{ruledtabular}
  \begin{tabular}{c c c c c}
    Thermal gradient& Time & Quenching & Phase & Ref. \\
    \hline
    280/210$^\circ$C & 1 week & No &$\alpha$-Bi$_4$I$_4$ & This work\\
    250/200$^\circ$C & 2 weeks & No & $\alpha$-Bi$_4$I$_4$& \cite{chen2018quantum}\\
    250/210$^\circ$C & 20-30 days &  No & $\beta$-Bi$_4$I$_4$& \cite{wang2021quantum}\\
    250/210$^\circ$C & 20 days & Not mentioned & $\beta$-Bi$_4$I$_4$ & \cite{pisoni_pressure_2017}\\
    285/188$^\circ$C & 3 days &Yes - Liquid nitrogen & $\beta$-Bi$_4$I$_4$& \cite{noguchi_weak_2019}\\
  \end{tabular}
  \end{ruledtabular}
  \label{tab:Grow}
\end{table*}

Let us then scrutinize the conditions under which the different crystals were grown by Chemical Vapour Transport, which is the method mainly employed to obtain Bi$_4$I$_4$. A summary of the growth conditions parameters was compiled and reported in table \ref{tab:Grow}: a combination of slightly different thermal gradients, time of growth and the quenching process was employed. Those growth protocols resulted in electrical resistivity of Bi$_4$I$_4$ as reported in Fig. \ref{fig:FigRho}. So far, it is not established which temperature-dependent features $\rho(T)$ should display if only determined by the pristine phases. An ideal 3D topological insulator is expected to show saturation in electrical resistivity at low temperatures due to the fully developed insulating bulk and protected metallic surface states. However, in the case of Bi$_4$I$_4$, the topological classification is still an open debate, and, thus, it is not yet established which specific signature can be expected for each phase and for each topological state.

%\st{In a simplified picture, we could expect the  topologically trivial insulator $\alpha$-Bi$_4$I$_4$ to display an insulating behaviour at low temperatures and, thus, to show a divergent resistivity at low temperatures. In the latter scenario, we could point to Noguchi et al.'s sample as the one that comes closer to the putative ideal behaviour at least for the $\alpha$ phase.}
%%\walber{As pointed out in Ref.~\onelinecite{liu2022gate}, in $\alpha$-phase the bulk and surface states are gapped }

We can observe that Noguchi et al.'s sample \cite{noguchi_weak_2019} displays the highest value of resistivity at low temperatures.  Among the available works, they employed the highest upper temperature of the growth gradient (285$^\circ$C) and the lowest bottom temperature of the same (188$^\circ$C), and finally, they quenched by using nitrogen. Interestingly, they used the shortest growth time (3 days). This observation might suggest a path to explore for further improvement in the growth process aiming at achieving a Bi$_4$I$_4$ with the lowest level of native defects, which, besides inducing unintentional doping, might also preclude access to a metastable $\beta$-Bi$_4$I$_4$. Future investigations on the role of doping and phase stability will be needed to identify any signature of a specific topological phase in transport properties.

%We close the discussion by observing that it is not possible yet to conclude whether the defects in the as-cast sample are impeding the low-temperature phase $\alpha$ from reaching the $\beta$ phase during the thermal treatment, or whether the defects formed in the $\beta$ phase at high temperatures are making the structure unstable. %and, thus, the metastable Bi$_4$I$_4$ quickly returns to $\alpha$.

Our results show that this instability might hinder not only a correct attribution of the structural phase in a metastable Bi$_4$I$_4$ obtained upon quenching and not further verified, but also can overlook the presence of mixed phases in as-cast samples maintained in the surrounding of the martensitic structural transition temperature $T_P \sim 300$~K.  Further theoretical calculations and temperature-resolved XRD experiments across the transition can shed light on these open questions.

%Last but not least, even in stable $\alpha$ and $\beta$ different types of native defects may impact the expected features of the pristine topological states. Further theoretical calculations and temperature-resolved XRD experiments across the transition can shed light on these open questions.

%%%%%%%%%%%%%%%%%%%%%%%%%%%%%%%%%%%%%%%%%%%%%%%%%%%%%%%%%%%%
%%%%%%%%%%%%%%%%%%%%%    DISCUSSION    %%%%%%%%%%%%%%%%%%%%%
%%%%%%%%%%%%%%%%%%%%%%%%%%%%%%%%%%%%%%%%%%%%%%%%%%%%%%%%%%%%

%%%%%%%%%%%%%%%%%%%%%%%%%%%%%%%%%%%%%%%%%%%%%%%%%%%%%%%%%%%%
%%%%%%%%%%%%%%%%%%%%%   CONCLUSIONS    %%%%%%%%%%%%%%%%%%%%%
%%%%%%%%%%%%%%%%%%%%%%%%%%%%%%%%%%%%%%%%%%%%%%%%%%%%%%%%%%%%

\section{Conclusions}

We carried out a structural investigation on as-cast $\alpha$-Bi$_4$I$_4$ and compared it with the putative non-trivial topological $\beta$ phase of Bi$_4$I$_4$ after quenching. Within our samples, we found a quick transition of the $\beta$-Bi$_4$I$_4$ towards the low temperature $\alpha$ phase upon room temperature exposure ($\sim 25$ \textdegree C). 
Our Rietveld refinements find iodine vacancies to be present after growth and to increase after quenching, relating to the variation of $\rho(T)$ detected in $\alpha$-Bi$_4$I$_4$.
A comparative analysis of the growth parameters and the electrical resistivity at low temperatures for the $\alpha$ phase suggests a correlation between the temperature gradient used during the growth process and the maximum saturation value of the electrical resistivity that is displayed at the lowest investigated temperature. 
In the case of the $\beta$ phase, we argue that the instability at low temperatures might be a consequence of the introduction of defects after quenching and their effect on the martensitic phase transformation. 

Finally, our findings indicate an absence of a clear fingerprint in the electrical resistivity of the non-trivial topological phase of the $\beta$-Bi$_4$I$_4$. Future investigations of the role of the different impurities on the electronic density of states and the impact on electrical conductivity can shed light on this aspect to further discern between features determined by the phase or by point defects.

\section*{Acknowledgments}
We thank Geetha Balakrishnan for insightful discussions. We acknowledge the ﬁnancial support of São Paulo Research Foundation (FAPESP), V.M., C.D.H, and JLJ under the grants 2018/19420-3, 2022/00992-2, 2018/08845-3; J.S.M. and L.R.F under the grants 2021/14322-6, 2021/08966-8; and also the support of CNPq (in particular Grant 402919/2021-1). W.H.B. and G.C. acknowledge FAPEMIG and the National Laboratory for Scientific Computing (LNCC/MCTI, Brazil) for providing HPC resources of the SDumont supercomputer, which have contributed to the research results, URL: http://sdumont.lncc.br. JLJ acknowledges CNPq Grant N. 310065/2021-6. We thank A. C. Franco of the Multi-user Laboratory of Crystallography at IF-USP for the support during the XRD measurements.

%\bibliography{refDH}

%apsrev4-2.bst 2019-01-14 (MD) hand-edited version of apsrev4-1.bst
%Control: key (0)
%Control: author (8) initials jnrlst
%Control: editor formatted (1) identically to author
%Control: production of article title (0) allowed
%Control: page (0) single
%Control: year (1) truncated
%Control: production of eprint (0) enabled
%

\end{document}